\documentclass[a4paper,prl,twocolumn,amsmath,amssymb,longbibliography]{revtex4-1}
\usepackage{bm}
\usepackage{graphicx,color}
\usepackage{soul}

\newcommand{\dd}{\mathrm{d}}

\newcommand{\pd}[2]{\frac{\partial #1}{\partial #2}}

\newcommand{\Int}[1]{\int\dd #1\;}
\newcommand{\IInt}[3]{\int_{#2}^{#3}\dd #1\;}
\newcommand{\OInt}[2]{\oint_{#2}\dd #1\;}

\renewcommand{\vec}[1]{\mathbf #1}

\newcommand{\al}{\alpha}

\newcommand{\eps}{\varepsilon}

\newcommand{\vhi}{\varphi}
\newcommand{\sig}{\sigma}
\newcommand{\om}{\omega}
\newcommand{\Om}{\Omega}

\newcommand{\id}{\mathbf 1}

\newcommand{\x}{\vec r}
\newcommand{\X}{\vec R}

\newcommand{\Dr}{D_\text{r}}
\newcommand{\tx}{\tau_\text{r}}
\newcommand{\meps}{\bm\eps}
\newcommand{\msig}{\bm\sigma}

\newcommand{\kT}{k_\text{B}T}
\newcommand{\ff}{\text{ff}}
\newcommand{\Lh}{\tfrac{L}{2}}
\newcommand{\nois}{\boldsymbol\xi}

\begin{document}

\title{Vorticity Determines the Force on Bodies Immersed in Active Fluids}

\author{Thomas Speck}
\author{Ashreya Jayaram}
\affiliation{Institut f\"ur Physik, Johannes Gutenberg-Universit\"at Mainz, Staudingerweg 7-9, 55128 Mainz, Germany}

\begin{abstract}
  When immersed into a fluid of active Brownian particles, passive bodies might start to undergo linear or angular directed motion depending on their shape. Here we exploit the divergence theorem to relate the forces responsible for this motion to the density and current induced by--but far away from--the body. In general, the force is composed of two contributions: due to the strength of the dipolar field component and due to particles leaving the boundary, generating a non-vanishing vorticity of the polarization. We derive and numerically corroborate results for periodic systems, which are fundamentally different from unbounded systems with forces that scale with the area of the system. We demonstrate that vorticity is localized close to the body and to points at which the local curvature changes, enabling the rational design of particle shapes with desired propulsion properties.
\end{abstract}

\maketitle


The defining feature of mesoscopic active matter~\cite{roadmap,bech16} is the persistent Brownian motion of its constituents, \emph{i.e.}, particles have a tendency to continue moving in the direction of previous displacements. The resulting motion is characterized as self-propulsion or, in a solvent, ``microswimming''. In contrast to passive diffusion, such an orientational persistence breaks detailed balance and thus requires steady dissipation. Some of the dissipated heat can be reclaimed as useful work and strategies on how to turn active matter into living and synthetic microengines have been proposed recently~\cite{vizs17,piet19}. Designing and optimizing such engines requires a comprehensive theoretical understanding of the forces generated in active fluids and suspensions~\cite{spec20}.

Asymmetry plays a crucial role for self-propulsion. Spherical colloidal particles can break their symmetry through different surface properties, most often in the form of Janus particles with two distinct hemispheres, one active and the other one inert~\cite{hows07,gole07}. Active but uniform particles might remain inert individually, but mixed with passive particles they aggregate into clusters that show directed linear and angular motion depending on the cluster's symmetry~\cite{soto14a,niu18a,lieb18}. To be distinguished from these modular microswimmers~\cite{niu18} are passive bodies placed into a suspension of, typically much smaller, active particles. Due to their directed motion, these particles get trapped and aggregate at the body's surface~\cite{fily14}, which induces a non-uniform density profile that is accompanied by an active stress and collective forces~\cite{spec20}. That passive but asymmetrically shaped bodies are subjected to directional forces has been demonstrated experimentally using microgears immersed into a bacterial bath~\cite{leon10,sokolov10,vizs17}, which undergo forward rotation on average. Related phenomena are current rectification~\cite{galajda07,wan08,mahmud09,sten16} and trapping~\cite{kaiser12,kumar19} of active particles due to single obstacles or arrays of fixed obstacles. Trapping and aggregation also occurs on walls confining an active fluid, giving rise to the notion of active (or swim) pressure, a phenomenon that has been under intense scrutiny~\cite{taka14,yan15,solo15,solo15a,ginot15,nikola16,speck16,junot17,fily18,duzgun18,solon18}.

Forces on (rigid) bodies immersed in active fluids have been studied numerically~\cite{angelani09,mallory14,ni15,small15,leite16,yamchi17}. Analytical studies have followed two strategies: While Yan and Brady focus on the density distribution within the interaction layer of trapped active particles close to the body~\cite{yan18}, Baek~\emph{et al.} relate the force to a dipolar algebraic decay of density and current in the far-field regime away from the body~\cite{baek18,granek20}. Here we connect both approaches through the divergence theorem, which relates the force due to a complicated density distribution in the interaction layer to the--potentially much simpler--solution for a free fluid far away from the body. We show that the force has two contributions, one from the far-field dipole and one from the vorticity of the polarization diffusing out of the interaction layer. While bodies in unbounded systems indeed appear as dipoles from a distance, this changes fundamentally in finite systems, where now the vorticity cancels the long-range dipolar field and sustains the force on the immersed body.


We study a ``dry'' fluid of non-interacting active particles moving with constant speed $v_0$ in two dimensions. The joint probability $\psi(\x,\vhi;t)$ obeys the evolution equation
\begin{equation}
  \label{eq:psi}
  \partial_t\psi = -\nabla\cdot[v_0\vec e\psi + \mu_0\vec F\psi - D_0\nabla\psi] + \frac{1}{\tx}\pd{^2\psi}{\vhi^2},
\end{equation}
where $\vec e\equiv(\cos\vhi,\sin\vhi)^T$ is the unit orientation that undergoes rotational diffusion with correlation time $\tx$. We assume that the translational diffusion coefficient $D_0=\kT\mu_0$ is related to the bare mobility $\mu_0$ through the temperature $T$ (with Boltzmann's constant $k_\text{B}$), which could be an effective temperature. The force $\vec F(\x)$ onto the active particles stems from walls and immersed objects and does not depend on the orientation $\vhi$.

We follow the standard route~\cite{saintillan15,spec20} and consider the hierarchy of moments with respect to the orientation $\vhi$. Integrating Eq.~\eqref{eq:psi} over $\vhi$ leads to the continuity equation $\partial_t\rho+\nabla\cdot\vec j=0$ with density $\rho(\x,t)\equiv\IInt{\vhi}{0}{2\pi}\psi$ and particle current
\begin{equation}
  \label{eq:j}
  \vec j = v_0\vec p + \mu_0\vec F\rho - D_0\nabla\rho.
\end{equation}
Here the polarization $\vec p(\x,t)\equiv\IInt{\vhi}{0}{2\pi}\vec e\psi$ enters. In the following, we drop the time dependence and consider the steady state with $\nabla\cdot\vec j=0$ everywhere. Multiplying Eq.~\eqref{eq:psi} by $\vec e$ followed again by integration over $\vhi$ now yields
\begin{equation}
  \label{eq:p}
  0 = -\nabla\cdot\left[\frac{v_0}{2}\rho\id+v_0\vec Q+\mu_0\vec F\vec p-D_0\nabla\vec p\right] - \frac{1}{\tx}\vec p,
\end{equation}
where $\vec Q$ is the nematic tensor. To close the hierachy, this tensor can be approximated as $\vec Q\approx-(v_0\tx/16)(\nabla\vec p)^{ST}$~\cite{bert06}. For the direct product, we write $(\vec{F}\vec{p})_{ij}=F_ip_j$ in cartesian coordinates and we employ the Einstein sum convention over repeated indices. The symmetric and traceless derivative reads $(\nabla\vec p)^{ST}_{ij}=\partial_ip_j+\partial_jp_i-(\nabla\cdot\vec p)\delta_{ij}$. Through Eq.~\eqref{eq:p}, the polarization $v_0\vec p=\mu_0\nabla\cdot\msig_\text{A}$ can be related to the divergence of a tensor
\begin{equation}
  \label{eq:sig:a}
  \msig_\text{A} = -v_0\tx\left[\frac{v_0}{2\mu_0}\rho\id+\frac{v_0}{\mu_0}\vec Q+\vec F\vec p-\kT(\nabla\vec p)^{ST}\right]
\end{equation}
that can be interpreted as an active stress. Note that we symmetrize the last term, which does not change the divergence. Replacing $v_0\vec p$ in Eq.~\eqref{eq:j}, we thus find the balance equation
\begin{equation}
  \label{eq:equ}
  \nabla\cdot\msig + \vec F\rho = \vec j/\mu_0
\end{equation}
with total stress tensor $\msig=-\kT\rho\id+\msig_\text{A}$. A uniform fluid has constant density $\rho=\rho_\infty$ and vanishing polarization, $\vec p=0$, for which the stress is isotropic.

We consider immersed objects (and possibly walls) with a finite interaction range, \emph{i.e.}, beyond a certain (microscopic) distance the force $\vec F=0$ vanishes and the active particles away from walls and bodies move freely. This defines disjoint regions, a ``free'' region in which $\vec F=0$ bounded by several regions with $\vec F\neq 0$ corresponding to each body within which the density $\rho$ of active particles declines sharply, defining a thin interaction layer at the surface of bodies. Due to their persistent motion, active particles are trapped at bodies and accumulate within the interaction layer generating an inhomogeneous density $\rho(\x)$ also in the free region. More importantly, such an inhomogeneous density is accompanied by a non-vanishing polarization and, consequently, a non-isotropic active stress Eq.~\eqref{eq:sig:a}.

The force $\vec F_1$ on a single body is minus the total force exerted on the fluid. One route is to estimate the density within the interaction layer~\cite{yan15a,duzgun18}. Here we follow a different strategy and, together with Eq.~\eqref{eq:equ}, exploit the divergence theorem to rewrite
\begin{equation}
  \label{eq:Fb}
  \vec F_1 = -\IInt{^2\x}{A}{} \vec F\rho = \OInt{l}{\partial A} \left[\vec n\cdot\msig - \frac{1}{\mu_0}(\vec n\cdot\vec j)\x\right]
\end{equation}
as a line integral along the closed curve $\partial A$ bounding the integration area $A$ (with normal vector $\vec n$ pointing outwards). For the second term we have used the identity $j_k=\partial_i(j_ix_k)$ with $\nabla\cdot\vec j=\partial_ij_i=0$. The integration area $A$ completely covers the body but is otherwise arbitrary (since $\vec F=0$ in the free region). The curve $\partial A$ lies completely within the force-free region, which reduces the determination of the local stress and current to a linear and homogeneous boundary-value problem away from the body.


\begin{figure}[t]
  \centering
  \includegraphics{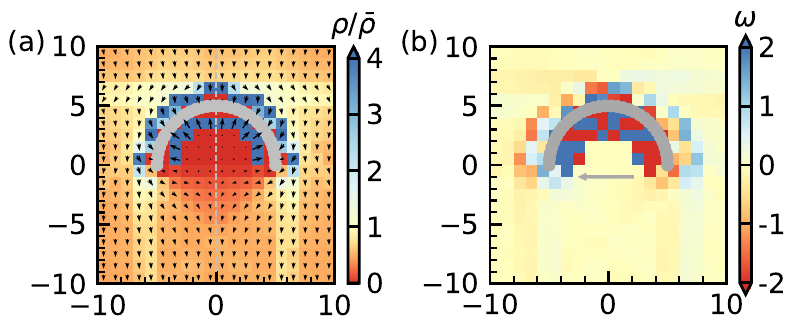}
  \caption{Line-symmetric boomerang-shaped body. (a)~Reduced density $\rho/\bar\rho$ (heat map) and polarization $\vec p$ (arrows) of active particles (with global density $\bar\rho=N/L^2=1.2$ and speed $v_0=80$). Shown is a close-up of the body, the total system is larger. The length of arrows is calculated as the logarithm of the local polarization magnitude. (b)~Vorticity map $\om$. Note the antisymmetric $\om(-x,y)=-\om(x,y)$ yielding a vanishing monopole moment $Q=0$ and a non-vanishing dipole moment $\vec P=P\vec e_x$ (arrow).}
  \label{fig:boom}
\end{figure}

To demonstrate the emergence of orientational order, we perform numerical simulations of $N$ non-interacting active particles surrounding a ``boomerang''-shaped body (semicircle with radius $R$) shown in Fig.~\ref{fig:boom}. We employ a square box with edge length $L$ and periodic boundary conditions (Supplemental Information~\cite{sm}). The shape is constructed as overlapping points repelling the active particles through the finite-range Weeks-Chandler-Andersen potential. Throughout, numerical results are presented in units of the potential length scale $a$ and time $a^2/D_0$. We calculate the local density and polarization on a regular grid plotted in Fig.~\ref{fig:boom}(a). We observe the aggregation of active particles directly at the surface of the body accompanied by a non-vanishing polarization due to the trapping of particles. But even away from the body there is a polarization of the active fluid with a pattern that is reminiscent of flow around a body. We stress that there is no alignment between particle orientations and this ordering is entirely due to the different retention times of particles at the top and bottom side of the boomerang.

To obtain the force on the boomerang in an infinite system it is sufficient to consider the decay of density and current far away from the body thanks to Eq.~\eqref{eq:Fb}. We again turn to Eq.~\eqref{eq:p}, which after eliminating the polarization through $v_0\vec p=D_0\nabla\rho+\vec j$ [Eq.~\eqref{eq:j}] becomes
\begin{equation}
  \label{eq:rho:j}
  -D_0(\nabla^2-\xi^{-2})\nabla\rho = (\nabla^2-\ell^{-2})\vec j
\end{equation}
in the free region. Here we have introduced two length scales,
\begin{equation}
  \ell \equiv \sqrt{D_0\tx}\left(1+\frac{v_0^2\tx}{16D_0}\right)^{1/2}
\end{equation}
and the decay length
\begin{equation}
  \label{eq:xi}
  \xi \equiv \ell \left(1+\frac{v_0^2\tx}{2D_0}\right)^{-1/2} \leqslant \ell
\end{equation}
with persistence length $v_0\tx$ of the directed motion. Assuming a body shape that does not generate currents ($\vec j=0$), we find $\nabla(\nabla^2\rho-\rho/\xi^2)=0$. To obtain the full density profile, this equation has to be solved for the densities prescribed at the boundaries of the free region. However, we can immediately infer that the excess density $\rho-\rho_\infty\sim e^{-r/\xi}$ decays exponentially far away from the body. Hence, we can push out the integration boundary $\partial A$ in Eq.~\eqref{eq:Fb} to distances where the density is uniform and, consequently, the stress is isotropic and the force thus zero.

Taking the divergence of Eq.~\eqref{eq:rho:j} yields for the density $(\nabla^2-\xi^{-2})\nabla^2\rho=0$ with now $\nabla^2\rho\sim e^{-r/\xi}\approx 0$ sufficiently far away from the body~\cite{yan18}. The far-field density profile is thus the solution of $\nabla^2\rho^\ff=0$ with current
\begin{equation}
  \label{eq:j:CR}
  \vec j^\ff = -D_\text{eff}\nabla\rho^\ff, \qquad 
  D_\text{eff} \equiv D_0(\ell/\xi)^2.
\end{equation}
An alternative route is to assume that the polarization decays slowly (on lengths much larger than $\ell$). In this limit, we can drop the second derivative in Eq.~\eqref{eq:p} to obtain $\vec p^\ff\approx-\frac{1}{2}v_0\tx\nabla\rho^\ff$ and thus Eq.~\eqref{eq:j:CR}.

The determination of the far-field current and density thus reduces to a problem that is well known from magnetostatics (with ``field'' $\vec j$ and scalar potential $\rho$). Let us consider the \emph{vorticity}
\begin{equation}
  \label{eq:om}
  \om(\x) \equiv \nabla\times\vec p \equiv \eps_{ij}\partial_ip_j = \partial_xp_y - \partial_yp_x
\end{equation}
of the polarization, where $\eps_{ij}$ is the Levi-Civita symbol with entries $\eps_{ii}=0$, $\eps_{12}=-\eps_{21}=1$. This vorticity has to be generated by the body, which is demonstrated for the boomerang in Fig.~\ref{fig:boom}(b). We find that at the ends of the arc vorticity is largest with opposite signs. Taking the curl of Eq.~\eqref{eq:rho:j}, the left hand side vanishes and we obtain the Helmholtz equation $\nabla^2\om-\om/\ell^2=0$ in the free region, corroborating the numerical observation that the vorticity decays fast (on the length $\ell$) and is localized close to the body. It would be exactly zero for currents of the form Eq.~\eqref{eq:j:CR} at variance with Fig.~\ref{fig:boom}(b).

The first two moments of the vorticity are the total ``charge'' $Q\equiv\Int{^2\x}\om=\OInt{\vec l}{}\cdot\vec p$ and the dipole moment $\vec P\equiv\Int{^2\x}\x\om$. In an unbounded system $Q=0$ (the boundary is at infinity) and there can be no angular current $\vec j_\theta\sim\vec e_\theta/r$. This should be intuitively clear given that rotational diffusion implies that without a density gradient the local polarization relaxes to zero, preventing such a rigid-body-like rotation of the fluid. The dipole moment
\begin{equation}
  \label{eq:P}
  \vec P(A) = \OInt{l}{\partial A}\left[(\vec n\times\vec p)\x-\frac{\mu_0}{v_0}\meps\cdot(\vec n\cdot\msig_\text{A})\right]
\end{equation}
can be expressed as a contour integral after performing an integration by parts and replacing $v_0\vec p=\mu_0\nabla\cdot\msig_\text{A}$ in the bulk term followed by the divergence theorem. Expanding the current $\vec j(\x)$ expressed through the Biot-Savart law, the far-field solution is given by the dipole field
\begin{equation}
  \label{eq:rho:j:ff}
  \vec j^\ff = \frac{1}{2\pi}\left[\frac{2(\vec m\cdot\x)\x}{r^4}-\frac{\vec m}{r^2}\right],
  \quad
  \rho^\ff = \rho_\infty + \frac{1}{D_\text{eff}}\frac{\vec m\cdot\x}{2\pi r^2}
\end{equation}
with current dipole moment $\vec m\equiv v_0\meps\cdot\vec P-\mu_0\vec F_1$ (Supplemental Information~\cite{sm}). It has two contributions, the first due to the vorticity of the polarization surrounding the body and the second due to the forces within the interaction layer, both contributing to $\Omega\equiv\nabla\times\vec j$. However, plugging the far-field solution Eq.~\eqref{eq:rho:j:ff} into Eq.~\eqref{eq:P} we find for an unbounded system $\vec P_\text{unb}=0$ independent of $A$ and $\vec F_1=-\vec m/\mu_0$ in agreement with Ref.~\citenum{baek18}. The same result is obtained plugging Eq.~\eqref{eq:rho:j:ff} into Eq.~\eqref{eq:Fb}~\cite{sm}. Note, however, that Fig.~\ref{fig:boom}(b) demonstrates that the vorticity dipole moment $\vec P$ does not vanish in the simulations, again indicating that the dipolar solution [Eq.~\eqref{eq:rho:j:ff}] is not applicable.


\begin{figure}[b!]
  \centering
  \includegraphics{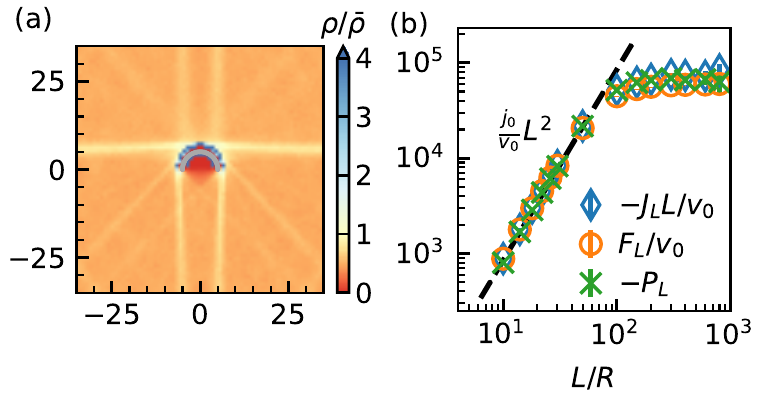}
  \caption{Finite-size behavior in periodic systems. (a)~Density profile $\rho$ for the same parameters as Fig.~\ref{fig:boom} but showing the full system. Note the strong departure from a dipolar field with ``streaks'' connecting through the periodic boundaries. (b)~Numerical current $-J_LL/v_0$, force $F_L/v_0$, and vorticity dipole moment $-P_L$ as a function of box size $L$. All quantities agree and scale as $L^2$ for small system sizes (dashed line with $j_0/v_0\simeq0.34$) before crossing over to a constant value $-P_\infty$ beyond $L^\ast\simeq50R$. Numerical data is shown for $v_0=160$.}
  \label{fig:fss}
\end{figure}

Computer simulations are necessarily performed in a finite simulation box employing periodic boundaries. Naively assuming a lattice of dipoles, within a shell of radii $r$ sufficiently far from the body \emph{and} boundary ($R\ll r\ll L$), to leading order the density is unmodified and given by the dipole Eq.~\eqref{eq:rho:j:ff}~\cite{sm}. However, the numerical density shown in Fig.~\ref{fig:fss}(a) computed for a system with $L/R=14$ deviates strongly from this expected profile. The current streamlines, instead of bending back as for the dipole, now connect through the periodic boundaries. This results in a total current
\begin{equation}
  J_L \equiv \IInt{x}{-L/2}{L/2} j_y(x,y) < 0
\end{equation}
through a cross section spanning the system, which has to be independent of $y$ since the divergence of the particle current is zero. As shown in Fig.~\ref{fig:fss}(b), the current $J_L=j_0L$ increases linearly for small cross sections $L$ and decays as $J_L\sim1/L$ for large $L$ (the scaling is discussed in more detail in the Supplemental Information~\cite{sm}).

In contrast to unbounded systems, from symmetry arguments alone we find that $\OInt{l}{\partial A}\vec n\cdot\msig=0$ in periodic systems, where the contour $\partial A$ is the square bounding the simulation box (Supplemental Information~\cite{sm}). As a corollary, the total polarization $\Int{^2\x}\vec p=(\mu_0/v_0)\OInt{l}{}\vec n\cdot\msig_\text{A}=0$ also vanishes, which is confirmed by the simulations. Hence, the polarization in the free region and within the interaction layer compensate each other. The force on the body
\begin{equation}
  \vec F_1 = -\frac{1}{\mu_0}\OInt{l}{\partial A}(\vec n\cdot\vec j)\x = F_L\vec e_y
\end{equation}
is now entirely determined by the size-dependent current with $F_L=-LJ_L/\mu_0$~\cite{sm}. This relation is confirmed numerically in Fig.~\ref{fig:fss}(b). In addition, we determine the vorticity dipole moment $\vec P=P_L\vec e_x$ through numerical integration of the vorticity field $\om$. Figure~\ref{fig:fss}(b) shows that the numerical values $-P_L$ agree with the force $\mu_0F_L/v_0$ as predicted theoretically from Eq.~\eqref{eq:P} with $v_0P_L=j_0L^2$~\cite{sm}.

The origin of the force on an immersed body in a periodic system is thus fundamentally different from the unbounded system: The strength of the far-field dipole $\vec m=v_0\meps\cdot\vec P-\mu_0\vec F_1=-(v_0P_L+\mu_0F_L)\vec e_y=0$ vanishes and the idea of a dipole lattice is not applicable. The vorticity $\om$ diffusing into the free region now determines the current and the force on the body. This force can become very large, scaling as $F_L\sim L^2$ before saturating beyond $L^\ast\sim\sqrt{\ell R}$ to a constant force $F_\infty\sim v_0^2$ that increases quadratically with the speed. This insight has practical ramifications for the design of engines and pumps as it implies an optimal spacing of obstacles with separation $L^\ast$ that maximizes the current and the force per obstacle. Moreover, it implies that inclusions cannot be modeled as dipoles in computer simulations.


\begin{figure}[t]
  \centering
  \includegraphics{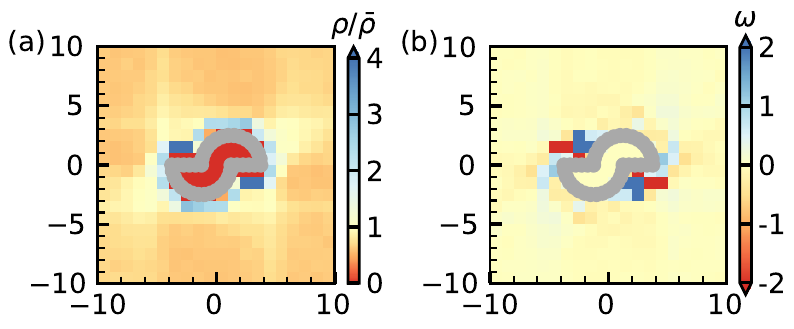}
  \caption{C$_2$-symmetric body. (a)~Density profile and (b)~vorticity map. Note the two dipoles forming at the shape's corners, which are responsible for the non-vanishing torque (while $Q=0$).}
  \label{fig:DD}
\end{figure}

Next, we perform simulations of the shape shown in Fig.~\ref{fig:DD}, which consists of two hemicircles cut and displaced along the $x$-axis. In agreement with its symmetry (no line symmetry, C$_2$-symmetric), we measure no linear force $\vec F_1=0$ but a non-vanishing torque $\tau_1$. The torque can be written
\begin{equation}
  \tau_1 = -\IInt{^2\x}{A}{} \x\times\vec F\rho = \IInt{^2\x}{A}{} \x\times[\nabla\cdot\msig-\vec j/\mu_0],
\end{equation}
where we have inserted the force balance Eq.~\eqref{eq:equ}. Again, we have an integration area $A$ that has to cover the body but is otherwise arbitrary. It is straightforward but somewhat tedious to rewrite this expression so that it only contains the current (Supplemental Information~\cite{sm}). Inserting the current multipole expansion, we find $\tau_1\propto v_0Q$. But also in a periodic system $Q=0$ since the polarization on the boundary $\partial A$ has to obey the periodic boundary conditions so that opposite edges cancel in the line integral $Q=\OInt{\vec l}{\partial A}\cdot\vec p=0$.

The fact that $Q=0$ seems to contradict the non-vanishing torque that we measure in the simulations. Inspecting the numerical vorticity map Fig.~\ref{fig:DD}(b), we find indeed $Q=0$ and also $\vec P=0$. A closer look, however, reveals that now two local dipoles are present at the left and right ``overhangs'' of the shape with $\vec P_1+\vec P_2=0$. In an unbounded system, decomposing the far-field current field into the contributions of dipoles at $\x_n$ each with moment $\vec m_n$, $\vec j(\x)=\sum_n\vec j^\ff_n(\x-\x_n)$, we find for the torque
\begin{equation}
  \tau_1 = -\frac{1}{\mu_0}\IInt{^2\x}{A}{} \sum_n(\x_n+\x)\times\vec j^\ff_n(\x) = \sum_n \x_n\times\vec F_n
\end{equation}
using that $\IInt{^2\x}{A}{}\x\times\vec j^\ff_n\sim Q_n=0$, and $\vec F_n$ is the force of dipole $n$ with dipole moment $\vec m_n$. In finite systems, from the simulations we again find a quadratic scaling $\tau_1\sim v_0L^2$ with system size $L<L^\ast$~\cite{sm}.


\begin{figure}[t]
  \includegraphics[width=\linewidth]{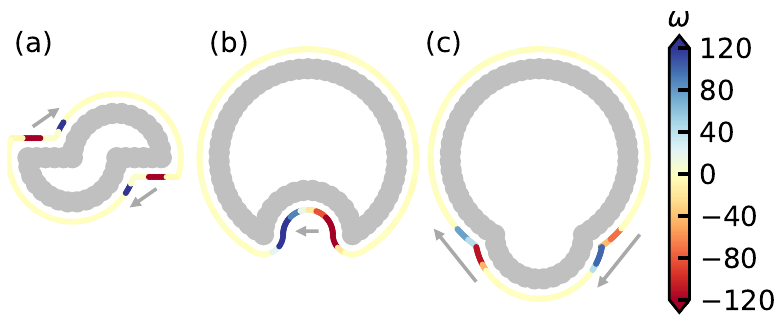}
  \caption{Vorticity $\om$ along the perimeter of three shapes: (a)~the shape from Fig.~\ref{fig:DD}, (b)~a lune (subtracting a smaller disc) and (c)~a bud (union with smaller disc). The vorticity is zero almost everywhere and localized to regions close to changes of the local curvature. Arrows indicate the local dipoles $\vec P_n$, which are oriented oppositely to the orientation of the perimeter.}
  \label{fig:vort}
\end{figure}

Figure~\ref{fig:vort} shows three shapes plotting the vorticity along the perimeter of each shape just outside the interaction layer. All shapes exhibit the same behavior: the vorticity $\om$ is strongly localized close to points where the local curvature changes and is zero elsewhere. At these points local dipoles $\vec P_n$ emerge, which determine whether the body experiences a linear force $\vec F_1=(v_0/\mu_0)\sum_n\meps\cdot\vec P_n$, a torque $\tau_1=-(v_0/\mu_0)\sum_n \x_n\cdot\vec P_n$, or both. This insight can be used to design shapes with the desired behavior.


To conclude, bodies immersed in an active fluid can induce currents depending on their shape, which act back on the body and cause linear and angular propulsion. The analogy of the underlying equation~\eqref{eq:rho:j} in an unbounded force-free region is with magnetostatics (not electrostatics), whereby the field is generated by a steady ``electric'' current density $\Om\vec e_z$ perpendicular to the plane of motion. While the force in an unbounded system is determined by the dipole contribution alone, passive inclusions are \emph{not} force dipoles and the analogy breaks down for periodic arrays and periodic boundary conditions. Now the force is sustained by the polarization vorticity leaving the interaction layer and enables giant forces and torques.

Here we have studied several simple shapes with symmetries causing either linear or angular propulsion. Our results can be tested in experiments, \emph{e.g.} in periodic arrays of rotors driven by bacteria~\cite{vizs17}. Directed forces might be harvested for the self-assembly of passive bodies through active suspensions. An intriguing question concerns the optimal shape that maximizes the vorticity dipole moment $\vec P$. Refs.~\citenum{kaiser12,kumar19} study the trapping of active particles in a wedge as a function of the opening angle. For our purposes we require a large accumulation together with a large exit current, which seems to be the case for intermediate angles $\sim\pi/2$ but needs to be investigated more carefully in future work.


\begin{acknowledgments}
  We acknowledge funding by the Deutsche Forschungsgemeinschaft (DFG) within collaborative research center TRR 146 (Grant No. 404840447) and the priority program SPP 1726 (Grant No. 254473714). Computations have been performed on the supercomputer MOGON II.
\end{acknowledgments}

%


\newpage

\section*{Supplemental Information}

\section*{Brownian dynamics simulations}

We study an $L\times L$ periodic system of $N=\bar\rho L^2$ non-interacting active particles with global number density $\bar\rho$. The boomerang-shaped inclusion is composed of equidistant points that lie on a semicircular arc of radius $R=5$. These points act as force centers and the $k$th point interacts with the $l$th active particle at a distance $r_{kl}$ through a shifted Weeks-Chandler-Andersen potential given by
\begin{equation*}
  u_{kl} = 
  \begin{cases}
    4\eps_0\left[\left(\frac{a}{r_{kl}}\right)^{12}-\left(\frac{a}{r_{kl}}\right)^6+\frac{1}{4}\right] & \text{if } r_{kl}\leq 2^{1/6}a\\
    0 & \text{otherwise}.
  \end{cases}
\end{equation*}
Here, $a$ is the potential's length scale and $\eps_0$ is the depth of the energy well. The body shown in Fig.~3 is similarly constructed with points placed on the contour of the shape obtained by displacing two semicircles of radius $R_\text{s}=2.5$ that share their diameters along the diameter. The evolution of the position $\x_k$ and orientation $\vhi_k$ of the $k$th active particle is governed by overdamped equations of motion:
\begin{equation}
  \label{eq:eom}
  \dot\x_k = v_0\vec e_k-\mu_0\nabla_k U+\sqrt{2D_0}\nois,\qquad \dot\vhi_k = \sqrt{2\Dr}\xi_\text{r}
\end{equation}
where $v_0$ is the propulsion speed, $D_0$ is the translational diffusion coefficient, $\Dr$ is the rotational diffusion coefficient and $U=\sum_{k,l}u_{kl}$ is the total potential energy due to interactions with the immersed passive body. The mobility $\mu_0=D_0/(\kT)$ where $\kT$ is the thermal energy. The components of $\nois$ and $\xi_\text{r}$ are drawn from a uniform distribution over $[-\sqrt{3},\sqrt{3}]$. We measure lengths in units of $a$, energy in units of $\kT$ and time in units of $a^2/D_0$. In non-dimensional units, we set $\eps_0=100$ and $\Dr=3$.

We integrate Eqs.~\eqref{eq:eom} with time step $10^{-5}$. The system is evolved for sufficiently long (at least until $\tau=100$ in simulation units) so that it reaches steady state following which we measure quantities of interest. All results are ensemble averages over 5 independent runs. For each run, we additionally compute a time average over at least 50 frames with a sampling frequency $\tau'=1$. The heatmaps shown in Figs.~1, 2, 3 in the main text are obtained by binning the simulation domain with a bin length of $a$ and computing concerned averages within each bin. Spatial derivatives are calculated on the grid using a central difference scheme that respects periodic boundaries. 
The force on the boomerang $\vec F_1$ is calculated by summing over the force on each point that composes it. Identically, the net torque $\tau_1$ is obtained by summing the two-dimensional cross product $\eps_{ij}x_iF_j$ over all points. Here, $x_i$ is the position of the point with respect to the center-of-mass of the body and $F_j$ is the force on the point.

\section*{Multipole expansion}

For completeness, we here provide the derivation of the far-field expressions for the current. The analog of the Biot–Savart law in two dimensions for the current yields (for clarity, we write the vector product in three dimensions)
\begin{equation*}
  \vec j(\x) = \frac{1}{2\pi}\Int{^2\x'}\frac{[\Om(\x')\vec e_z]\times(\x-\x')}{|\x-\x'|^2}
\end{equation*}
with curl
\begin{align*}
  \Om \equiv \nabla\times\vec j &= \nabla\times(v_0\vec p+\mu_0\vec F\rho-D_0\nabla\rho) \\ &= v_0\om + \mu_0\nabla\times(\vec F\rho)
\end{align*}
and $\om\equiv\nabla\times\vec p$. The first two moments are $\tilde Q\equiv\Int{^2\x}\Om$ and $\tilde{\vec P}\equiv\Int{^2\x}\x\Om$, whereby $\tilde Q=0$ due to the same reasons as discussed in the main text for $Q$.

Employing the Levi-Civita symbol $\eps_{ij}$, we find in cartesian coordinates
\begin{equation}
  \label{eq:si:j}
  j_i(\x) = -\frac{1}{2\pi}\eps_{ij}\Int{^2\x'}\frac{x_j-x_j'}{|\x-\x'|^2}\Om(\x').
\end{equation}
The Taylor expansion for small $\x'$ reads
\begin{align}
  \frac{x_j-x_j'}{|\x-\x'|^2} &\approx \frac{x_j}{r^2} + \left(\pd{}{x_k}\frac{x_j-x_j'}{|\x-\x'|^2}\right)_{\x'=0}x_k' \\
  \label{eq:taylor}
  &= \frac{x_j}{r^2} - \left(\pd{}{x_k}\frac{x_j}{r^2}\right)x_k'.
\end{align}
Plugging this expansion back into the current [Eq.~\eqref{eq:si:j}] leads to
\begin{equation*}
  j_i^\ff = \frac{1}{2\pi}\eps_{ij}\left(\pd{}{x_k}\frac{x_j}{r^2}\right)\tilde P_k = \frac{1}{2\pi}\eps_{ij}\pd{}{x_j}\frac{x_k\tilde P_k}{r^2},
\end{equation*}
where in the second step we have exchanged indices using the fact that the expression in brackets is symmetric. Performing the derivative and setting
\begin{align*}
  m_i \equiv \eps_{ij}\tilde P_j &= v_0\eps_{ij}P_j + \mu_0\eps_{ij}\Int{^2\x}x_j\nabla\times(\vec F\rho) \\
  &= v_0\eps_{ij}P_j - \mu_0\eps_{ij}\eps_{jl}\Int{^2\x}F_l\rho
\end{align*}
leads to the result $\vec m\equiv v_0\meps\cdot\vec P-\mu_0\vec F_1$ given in the main text, whereby the boundary of the integral lies within the free region and thus vanishes.

\section*{Unbounded system: Far-field regime}

\subsection*{Dipole moment}

The vorticity dipole moment reads
\begin{align*}
  P_k(A) &= \IInt{^2\x}{A}{} x_k\eps_{ij}\partial_ip_j \\ &= \OInt{\ell}{\partial A}x_k\eps_{ij}n_ip_j - \eps_{kj}\IInt{^2\x}{A}{} p_j
\end{align*}
after integration by parts. Eliminating the polarization in the second term, the dipole moment over a finite area $A$ can be expressed as the contour integral
\begin{align*}
  P_k(A) &= \OInt{\ell}{\partial A}\left[x_k(\vec n\times\vec p)-\frac{\mu_0}{v_0}\eps_{kj}n_i\sig^\text{A}_{ij}\right] \\
  &= -\frac{1}{2}v_0\tx\OInt{\ell}{\partial A}\left[x_k(\vec n\times\nabla\rho^\ff)-\eps_{kj}n_j\delta\rho^\ff\right],
\end{align*}
where in the second step we have inserted the far-field solution for polarization, $\vec p^\ff=-\frac{1}{2}v_0\tx\nabla\rho^\ff$, and active stress, $\mu_0\msig_\text{A}=-\frac{1}{2}v_0^2\tx\rho\id$ [the polarization gradient does not contribute, cf. Eq.~\eqref{eq:grad}]. Going to a circular boundary with radius $r$ and $\vec n=\vec e_r$, we find for an unbounded system
\begin{equation*}
  \vec P_\text{unb} = -\frac{1}{2}v_0\tx\OInt{\ell}{\partial A}\left[ \pd{\rho^\ff}{\theta}\vec e_r+\delta\rho^\ff\vec e_\theta \right] = 0
\end{equation*}
using $\meps\cdot\vec e_r=-\vec e_\theta$. Both terms cancel since
\begin{equation}
  \IInt{\theta}{0}{2\pi} (\cos\theta\vec e_r+\sin\theta\vec e_\theta) = 0.
  \label{eq:circ}
\end{equation}

\subsection*{Force}

The total stress tensor can be written
\begin{equation}
  \label{eq:sig}
  \mu_0\msig = -D_\text{eff}\rho\id + v_0\ell^2(\nabla\vec p)^{ST}
\end{equation}
with length $\ell$ and effective diffusion coefficient $D_\text{eff}$ given in the main text. In the far-field regime, $\vec p^\ff=-\frac{1}{2}v_0\tx\nabla\rho^\ff$ with density
\begin{equation*}
  \rho^\ff = \rho_\infty + \frac{1}{D_\text{eff}}\frac{\vec m\cdot\x}{2\pi r^2}.
\end{equation*}
We need the expression
\begin{multline}
  \vec e_r\cdot(\nabla\nabla\rho)^{ST} = \left(2\pd{^2\rho}{r^2}-\nabla^2\rho\right)\vec e_r \\ + \frac{2}{r}\left(\pd{^2\rho}{r\partial\theta}-\frac{1}{r}\pd{\rho}{\theta}\right)\vec e_\theta.
  \label{eq:rho:2nd}
\end{multline}
Inserting $\rho^\ff$, we obtain
\begin{equation*}
  \vec e_r\cdot(\nabla\nabla\rho^\ff)^{ST} = \frac{m}{D_\text{eff}}\frac{2}{\pi r^3}(\cos\theta\vec e_r+\sin\theta\vec e_\theta).
\end{equation*}
Integrating along a circle this contribution vanishes [cf. Eq.~\eqref{eq:circ}] and thus
\begin{equation}
  \OInt{\ell}{\partial A} \vec e_r\cdot(\nabla v_0\vec p)^{ST} = 0
  \label{eq:grad}
\end{equation}
vanishes along a closed circular contour $\partial A$. The calculation of the force in the far-field regime then reduces to
\begin{equation*}
  \mu_0\vec F_1 = -D_\text{eff}\OInt{\ell}{\partial A}(\delta\rho^\ff-\x\cdot\nabla\rho^\ff)\vec e_r = -\vec m,
\end{equation*}
to which stress and current both contribute $-\vec m/2$ each.

\section*{Periodic boundaries}

\subsection*{Density profile for a dipole lattice}

Employing periodic boundary conditions is equivalent to periodically replicating the system. The density at $\x$ due to assuming a square lattice of dipoles (each with dipole moment $\vec m$) becomes
\begin{equation*}
  \delta\rho^\ff_L(\x) = \sum_n\delta\rho^\ff(\x-\X_n) = \frac{1}{D_\text{eff}}\sum_n\frac{\vec m\cdot(\x-\X_n)}{2\pi|\x-\X_n|^2},
\end{equation*}
where $n$ sums over lattice vectors $\X_n$ with $\X_0=0$ the origin. We again employ the Taylor expansion Eq.~\eqref{eq:taylor} but now for small $|\x|\ll|\X_n|$,
\begin{equation*}
  \frac{x_i-X_i}{|\x-\X|^2} \approx -\frac{X_i}{\X^2} + \left(\pd{}{X_j}\frac{X_i}{\X^2}\right)x_j,
\end{equation*}
and thus
\begin{multline*}
  \delta\rho^\ff_L(\x) = \delta\rho^\ff(\x) + \\ \frac{1}{2\pi D_\text{eff}}\sum_{n\neq 0}\left[\frac{\vec m\cdot(\x-\X_n)}{\X_n^2}-\vec m\cdot\frac{2\X_n\X_n}{\X_n^4}\cdot\x\right].
\end{multline*}
There are four lattice vectors with the same length $|\X_n|=kL$, two of which sum to zero. Hence,
\begin{gather*}
  \sum_{n\neq 0}\frac{1}{\X_n^2} = \frac{4}{L^2}\sum_{k=1}^\infty\frac{1}{k^2} = \frac{4}{L^2}\frac{\pi^2}{6}, \qquad \sum_{n\neq 0}\frac{\X_n}{\X_n^2} = 0, \\
  \sum_{n\neq 0}\frac{\X_n\X_n}{\X_n^4} = \frac{2}{L^2}\id\sum_{k=1}^\infty\frac{1}{k^2} = \frac{\pi^2}{3L^2}\id
\end{gather*}
and the first and third term cancel; to first order of the Taylor expansion there is no correction to the density. Also to second order there is no correction since all terms at this order contain an odd number of lattice vectors.

\subsection*{Vanishing stress contour integral}



For the terms involving the current, we need the integrated current
\begin{equation*}
  J_L = \IInt{x}{-\Lh}{+\Lh} j_y(x,y).
\end{equation*}
With $\nabla\cdot\vec j=0$ and $j_x(\pm\Lh,y)=0$ we have
\begin{equation*}
  \IInt{x}{-\Lh}{+\Lh} \nabla\cdot\vec j = \IInt{x}{-\Lh}{+\Lh} (\partial_xj_x+\partial_yj_y) = \partial_y J_L = 0,
\end{equation*}
which implies that $J_L$ is uniform throughout the system.

\begin{figure}[b!]
  \centering
  \includegraphics{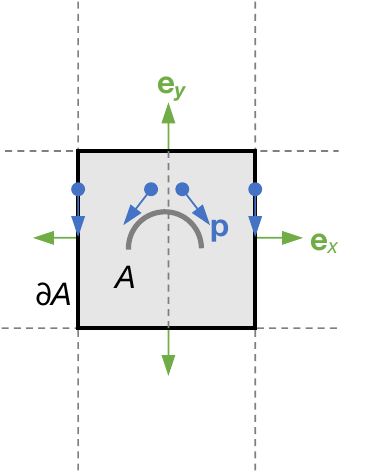}
  \caption{Sketch of the periodic system with area $A$ and bounded by the square $\partial A$ with normal vectors (green arrows). Sketched are the polarization $\vec p$ inside and at the left and right boundary (blue arrows).}
  \label{fig:periodic}
\end{figure}

We now investigate the stress Eq.~\eqref{eq:sig} for a line-symmetric body at the origin, cf. Fig.~\ref{fig:periodic}. This implies the symmetries
\begin{gather*}
  \rho(-x,y) = \rho(x,y), \\ p_x(-x,y) = -p_x(x,y), \\ p_y(-x,y) = p_y(x,y).
\end{gather*}
We integrate along the square contour $\partial A$ bounding the finite system with
\begin{equation*}
  \vec n\cdot\mu_0\msig = -D_\text{eff}\rho\vec n + v_0\ell^2\left[\nabla(\vec n\cdot\vec p)-\meps\cdot\nabla(\vec n\times\vec p)\right]
\end{equation*}
and piecewise constant normal vector $\vec n$. The first contribution from $\rho\vec n$ vanishes since normal vectors from left/right and top/bottom cancel each other (periodic density).

For the second contribution, let us look at the top boundary at $y=\Lh$ with $\vec n=\vec e_y$ leading to
\begin{equation*}
  v_0\IInt{x}{-\Lh}{+\Lh} [\nabla p_y+\meps\cdot\nabla p_x] = \partial_y\IInt{x}{-\Lh}{+\Lh}v_0\vec p = G\vec e_y,
\end{equation*}
where we used the periodicity of the polarization. In the second step, the $x$-component is zero because due to the symmetry of the body, the polarization is antisymmetric, $p_x(-x,y)=-p_x(x,y)$. For the $y$-component, we insert $v_0p_y=D_0\partial_y\rho+j_y$ using $\partial_yJ_L=0$ and define
\begin{equation*}
  G \equiv D_0\IInt{x}{-\Lh}{+\Lh} \partial_y^2\rho|_{y=\Lh}.
\end{equation*}
The bottom boundary has $\vec n=-\vec e_y$ leading to the same $G$ but now with the opposite sign so that these two contributions cancel.

For the right boundary at $x=\Lh$ with $\vec n=\vec e_x$ we obtain
\begin{equation*}
  v_0\IInt{y}{-\Lh}{+\Lh} [\nabla p_x-\meps\cdot\nabla p_y] = \partial_x\IInt{y}{-\Lh}{+\Lh}v_0\vec p = 0.
\end{equation*}
Now $p_x(\pm\Lh,y)=0$ and the symmetry $p_y(-x,y)=p_y(x,y)$ implies that the derivative $\partial_xp_y$ vanishes. The same holds for the left boundary at $x=-\Lh$ and thus
\begin{equation*}
  \OInt{\ell}{\partial A} \vec n\cdot\msig = 0.
\end{equation*}

\subsection*{Force, current, and dipole moment}

The direct contribution of the current to the force reads
\begin{multline*}
  \OInt{l}{\partial A}(\vec n\cdot\vec j)\x = \\
  \IInt{x}{-\Lh}{+\Lh} \left[j_y(x,-\Lh)\Lh+j_y(x,+\Lh)\Lh\right]\vec e_y = LJ_L\vec e_y
\end{multline*}
leading to the force $\vec F_1=F_L\vec e_y$ with $F_L=-LJ_L/\mu_0$. The $x$-component of the force vanishes because $j_y(-x)=j_y(x)$ and $j_y(x)x$ is odd.

Using that the total polarization vanishes, the dipole moment can be written
\begin{equation*}
  \vec P = \OInt{\ell}{\partial A} (\vec n\times\vec p)\x.
\end{equation*}
Inserting $v_0\vec p=D_0\nabla\rho+\vec j$, the density contribution vanishes and the only contribution left is
\begin{equation*}
  v_0\vec P = \IInt{y}{-\Lh}{\Lh} \left[j_y(-\Lh,y)\Lh+j_y(+\Lh,y)\Lh\right]\vec e_x = L^2j_0\vec e_x.
\end{equation*}
Since on the boundary $j_x=0$, divergence-free $\partial_yj_y=0$ requires $j_y(\pm\Lh,y)=j_0$ to be constant.

\subsection{Scaling of current}

\begin{figure}[t]
  \centering
  \includegraphics{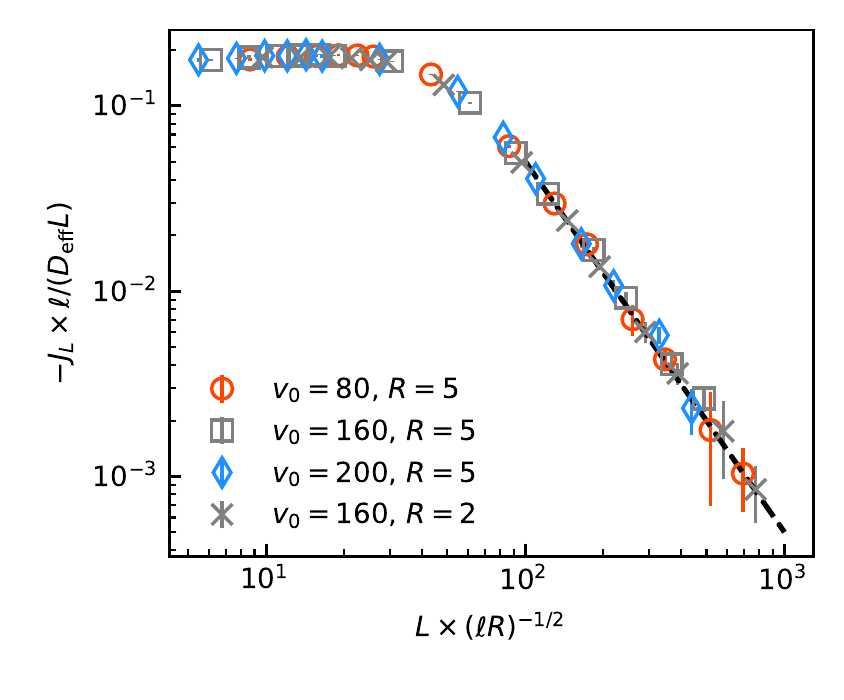}
  \caption{Integrated current $J_L$ for the boomerang as a function of the scaled system size $L/\sqrt{\ell R}$ for several speeds $v_0$ and two radii $R$. The dashed line shows $f\sim x^{-2}$.}
  \label{fig:si:fss}
\end{figure}

The current through the system is generated by the density difference of the accumulated active particles at the body. Eq.~(10) in the main text suggests
\begin{equation*}
  J_L = -D_\text{eff}\bar\rho\al_L
\end{equation*}
with dimensionless gradient $\al_L>0$ across the boomerang with radius $R$. For small $L$, increasing $L$ adds streamlines connecting through the periodic boundaries and increases the current, $\al_L\sim L/\ell$, with decay length $\ell$ governing the current. For large $L$, we expect $\al_L\sim R/L$ to be self-similar. Assuming that $\al_L$ depends on the system parameters only through the dimensionless combination $L/\sqrt{\ell R}$, the current through the cross section can be written
\begin{equation*}
  J_L = -\bar\rho D_\text{eff}\frac{L}{\ell}f(L/\sqrt{\ell R}) \sim \bar\rho D_\text{eff}
  \begin{cases}
    L/\ell & (L\ll L^\ast) \\ R/L & (L\gg L^\ast)
  \end{cases}
\end{equation*}
with scaling function $f(x)$, which is constant for small $x$ and decays as $f(x)\sim x^{-2}$ for large $x$. In Fig.~\ref{fig:si:fss}, we demonstrate that the numerical current as a function of $L/\sqrt{\ell R}$ indeed collapses onto a single curve $f(x)$ for several speeds $v_0$ and sizes $R$ of the obstacle. The crossover length $L^\ast$ scales as $L^\ast=x^\ast\sqrt{\ell R}$ with $x^\ast\simeq 50$.

Note that for large speeds $v_0\gg 4\sqrt{D_0/\tx}$, we have
\begin{equation*}
  \ell \approx \frac{v_0\tx}{4}, \qquad D_\text{eff} \approx \frac{1}{2}v_0^2\tx,
\end{equation*}
whereas $\xi\approx\sqrt{D_0\tx/8}$ becomes independent of the speed. For the force, we thus find
\begin{equation*}
  \mu_0F_L = -LJ_L \sim \bar\rho \begin{cases}
    v_0L^2 & (L\ll L^\ast) \\ v_0^2\tx R & (L\gg L^\ast)
  \end{cases}
\end{equation*}
in agreement with Figure 2(b) in the main text. This scaling is different from the one reported by Mallory \emph{et al.} [PRE \textbf{90}, 032309 (2014)] for a bath of underdamped self-propelled particles.

\section*{Torque}

\subsection*{Unbounded system}

Inserting the force-balance, the torque can be written as the integral
\begin{equation*}
  \tau_1 = \IInt{^2\x}{A}{} \eps_{ij}x_i[\partial_k\sig_{kj}-j_j/\mu_0].
\end{equation*}
The first term can be manipulated into
\begin{align*}
  \eps_{ij}x_i\partial_k\sig_{kj} &= \eps_{ij}[\partial_k(x_i\sig_{kj})-\sig_{ij}] \\ &= \eps_{ij}\partial_k(x_i\sig_{kj})
\end{align*}
using that the stress is symmetric, $\sig_{ij}=\sig_{ji}$. Exploiting the divergence theorem, we obtain
\begin{equation*}
  \tau_1 = \oint_{\partial A}\dd l\; \eps_{ij}x_in_k\sig_{kj} - \frac{1}{\mu_0} \IInt{^2\x}{A}{}\x\times\vec j.
\end{equation*}
The next step is to insert the stress choosing a circular boundary $\partial A$ with radius $r$ and normal vector $\vec n=\vec e_r$. It is easy to see that the isotropic part cancels since on the boundary $\x\times\vec n=0$. For the second derivatives of the density $\rho(r,\theta)$ we employ Eq.~\eqref{eq:rho:2nd}.
Taking the vector product $\x\times$ ($=\eps_{ij}x_i$) only the second term survives with $\x\times\vec e_\theta=r$. However, integrating this term along a closed circular boundary yields zero since the density necessarily is periodic with respect to $\theta$,
\begin{equation*}
  \IInt{\theta}{0}{2\pi} \pd{\rho}{\theta} = 0.
\end{equation*}
This leaves us with a torque [using $\eps_{ij}x_in_k\partial_kj_j=n_k\partial_k(\eps_{ij}x_ij_j)-\eps_{ij}n_ij_j$]
\begin{multline*}
  \tau_1 = \frac{\ell^2}{\mu_0}\oint_{\partial A}\dd l\; [\vec n\cdot\nabla(\x\times\vec j)+(\x\times\nabla)\vec n\cdot\vec j-\vec n\times\vec j] \\ - \frac{1}{\mu_0} \IInt{^2\x}{A}{}\x\times\vec j.
\end{multline*}
The second term involves $\partial_\theta j_r$ and vanishes after integration. Sticking with the circular boundary, the other three terms involve the integral
\begin{equation*}
  \IInt{\theta}{0}{2\pi} \x\times\vec j = \IInt{\theta}{0}{2\pi} rj_\theta = v_0Q
\end{equation*}
and thus $\tau_1\propto v_0Q$.

\subsection*{Periodic system}

\begin{figure}[b!]
  \centering
  \includegraphics{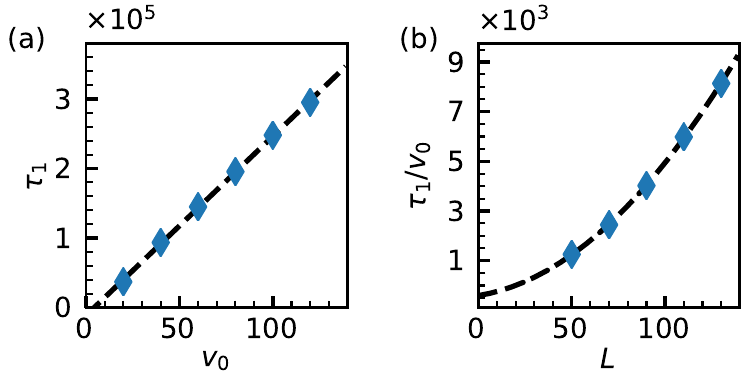}
  \caption{Numerical results for the torque on the C$_2$-symmetric body shown in Fig.~3 of the main text. Torque $\tau_1$ (a)~as a function of propulsion speed $v_0$ (the dashed line is a linear fit) and (b)~as a function of system size $L$ (the dashed line is a quadratic fit).}
  \label{fig:torque}
\end{figure}

Figure~\ref{fig:torque} shows numerical results for the torque exerted on the body discussed in the main page. As for translational forces, we find that initially $\tau_1\sim v_0L^2$ in finite systems with periodic boundaries.
  
\end{document}